\documentclass[a4paper,11pt]{article}

\usepackage{latexsym}
\usepackage{amsfonts}
\usepackage{color}
\usepackage{bbm}
\usepackage{amsmath}
\usepackage{amssymb}
\usepackage{tabularx}
\usepackage{chicagoz}
\usepackage{theorem}
\usepackage{pstricks,pst-plot}
\usepackage{graphicx}
\usepackage{epsfig}
\usepackage{lscape}
\usepackage{rotating}
\usepackage{float}
\usepackage{cite}
\usepackage{caption}
\usepackage{tabulary}
\usepackage{array}

\newcolumntype{L}[1]{>{\raggedright\let\newline\\\arraybackslash\hspace{0pt}}m{#1}}
\newcolumntype{C}[1]{>{\centering\let\newline\\\arraybackslash\hspace{0pt}}m{#1}}

\author{Eduardo Ramos-P\'erez$^{(1)}$,\\ Pablo J. Alonso-Gonz\'alez$^{(2)}$, Jos\'e Javier N\'u\~nez-Vel\'azquez$^{(2)}$ }
\title{Forecasting volatility with a stacked model based on a hybridized Artificial Neural Network}

\date{\footnotesize\textit{(1) Ph D Student (Economics and Management Program). Universidad de Alcal\'a.\\
	(2) Economics Department. Universidad de Alcal\'a.}
\thanks{\hskip -1.8em
   Authors' address: (1)\&(2) :Economics Department, Universidad de Alcal\'a, Plaza de la Victoria 2, 28802 Alcal\'a de Henares, Spain.
   E--mails: P.J. Alonso-Gonz\'alez, \texttt{pablo.alonsog@uah.es}, J.J. N\'u\~nez, \texttt{josej.nunez@uah.es}, E. Ramos,  \texttt{ramos.perez.e@gmail.com}}
\thanks{\hskip -1.8em Corresponding author: P.\, Alonso;\,\, Date: \today. This manuscript version is made available under the CC-BY-NC-ND 4.0 license http://creativecommons.org/licenses/by-nc-nd/4.0/}
}

\begin{document}
\maketitle
\begin{abstract}
\noindent
An appropriate calibration and forecasting of volatility and market risk are some of the main challenges faced by companies that have to manage the uncertainty inherent to their investments or funding operations such as banks, pension funds or insurance companies. This has become even more evident after the 2007-2008 Financial Crisis, when the forecasting models assessing the market risk and volatility failed. Since then, a significant number of theoretical developments and methodologies have appeared to improve the accuracy of the volatility forecasts and market risk assessments. Following this line of thinking, this paper introduces a model based on using a set of Machine Learning techniques, such as Gradient Descent Boosting, Random Forest, Support Vector Machine and Artificial Neural Network, where those algorithms are stacked to predict S\&P500 volatility. The results suggest that our construction outperforms other habitual models on the ability to forecast the level of volatility, leading to a more accurate assessment of the market risk.
\end{abstract}
{\small
}

\section{Introduction}
\label{introd}
During the Financial Crisis of 2007-2008, unexpected falls in stock prices resulted in significant losses for individual investors and financial institutions. Since then, new regulations have entered in force in order to ensure the correctness of the market risk assessment provided by financial institutions and to allow individual market participants to be aware of the risk linked to financial products. As volatility is an indicator of the uncertainty associated with the asset profitability (\citeNP{hul_2015} and \citeNP{rr_2015}), this variable tends to play a key role within the risk models. In fact, events like the bankruptcy of LTCM in 1998 (\citeNP{low_2000}), the dotcom crash in 2001 (\shortciteNP{agy_2010}) or, more recently, the aforementioned Financial Crisis of 2007-2008 were not foreseen by most of the risk models due to inaccurate estimates produced by the volatility forecasting models. It is worth mentioning that, as volatility is not directly observed, before estimating any statistical model it is necessary to select a volatility proxy (\citeNP{pg_2003}). In the following paragraphs, the proposed methodology and main families of volatility forecasting models (GARCH, Stochastic and Machine Learning) are presented.\\

First of all, GARCH models are introduced as this family of models is probably the most widely used in the literature due to its ability to fit the volatility clustering (\citeNP{man_1963}) empirically observed in financial time series. This auto-regressive approach and its generalization were developed by \citeN{eng_1982} and \citeN{Bollerslev_1986} respectively. Classical GARCH models were discovered to be too rigid for fitting returns series, especially over a long time span, because the estimated persistence of conditional variances is close to one (\shortciteNP{Bauwens_2012}). Therefore, more flexible GARCH models were developed in order to overcome this problem. \citeN{el_1999} suggested a two equation model where each of them represents long-run and short-run components of volatility, respectively. Mixed-normal GARCH (\shortciteNP{hmp_2004a}) is a second way to deal with this problem. This kind of model allows to choose amongst several regimes in each instant of time t. The drawback of this methodology is that it assumes that the variables used to decide amongst regimes are all independent over time. To overcome this problem, \shortciteN{hmp_2004b} proposed a Markov-switching model where the parameters of a GARCH model change according to a Markov process. An extension of this kind of model can be found in \citeN{hp_2012}. Before concluding with the GARCH models, it is important to mention that volatility can behave differently depending on the trend of the market: bullish or bearish. To fit this behaviour, \citeN{Nelson_1991} developed the EGARCH model that allows the sign and the volume of previous values to have separate impacts on the volatility forecasts. In addition to the EGARCH model, \shortciteN{gjr_1993} proposed the GJR-GARCH to replicate the aforementioned behaviour. Other developments within this family can be found in \citeN{ek_1995} with their BEKK model, the factor model (\shortciteNP{enr_1990}), the Constant Conditional Correlation model (\citeNP{bol_1990}), the time-varying correlation model (\citeNP{tt_2002}), the dynamic correlation model (\citeNP{eng_2002}) or the multivariate GARCH approach proposed by \citeN{ke_1982} and \shortciteN{egk_1984} and its financial implementation by \shortciteN{bew_1988}. More recently, \shortciteN{zzl_2018} have proposed a first order zero drift GARCH (ZD-GARCH) to study heteroscedasticity and conditional heteroscedasticity together.\\

The second family is composed of those models which assume that the volatility is driven by its own stochastic process. This approach was introduced by \citeN{tay_1982} as an Euler approximation of the underlying diffusion model. Assuming that stock prices follow a Brownian motion, \citeN{Heston_1993} derived a model where the volatility follows an Ornstein-Uhlenbeck process. To derive the parameters of the Heston Model, two different strategies have been adopted in the literature: moment or simulation. For the first one, the Generalized Method of Moments was proposed by \citeN{mt_1990} and \citeN{as_1999}, while the simulation approach has been used by \citeN{dan_2004}, \citeN{dk_1997}, \citeN{br_2004} or \citeN{and_2009}, amongst others.\\

The last family presented is Machine Learning, which comprises a set of techniques used to analyse the future evolution of stock prices and volatility. These algorithms try to learn automatically and recognize patterns in a large amount of data (\shortciteNP{kvf_2010}). It is worth mentioning that the fitting of these algorithms is quite sensitive to the forecasting time-frame and the selected input variables. \shortciteN{amm_2005} and \shortciteN{fag_2009} suggest using one day as a time-frame and lagged or technical indicators as input variables for the Machine Learning algorithms. Stock prices, volatilities and portfolio selection have been analysed using different methodologies based on Machine Learning, such as Support Vector Machine (\shortciteNP{gsb_2001}), hidden Markov models (\citeNP{gd_2012} and \shortciteNP{dias_2019}) or Artificial Neural Networks (ANN) (\citeNP{hi_2002}). These last authors showed that volatility forecasts made by an ANN outperform the implied volatility derived from Barone-Adesi and Whaley options models. Additionally, ANNs have been applied successfully to other financial series different from volatility and stock prices: bond rates (\citeNP{sx_2001}) and bank failures (\shortciteNP{hlw_1994}). Deep learning (\shortciteNP{lcun_2015}) is a framework closely related with ANN which has been employed for predicting the evolution of Korean stock market index (\shortciteNP{chp_2017}).\\

Despite the high performance of ANN, predictions derived from the use of this algorithm could be inaccurate when stock prices move sharply (\citeNP{py_2014}). To overcome this problem, ANN were combined with other statistical models (\shortciteNP{kfm_2014}) creating the so called hybrid models. Hybridization can be defined as an approach in which several models are merged to form a new enhanced model in order to produce better forecasting results. Therefore, a hybrid model is a combination of the artificial intelligence techniques with some components of the traditional forecasting models (like the ones presented within the GARCH family). Examples of this approach are discussed in \citeN{Roh_2006}, \shortciteN{hsz_2012}, \shortciteN{lu_2016} \shortciteN{m_2014} or \shortciteN{kfm_2014}, where different outputs from a GARCH-based model are used as inputs in an ANN. A more general picture of this type of hybrid models is provided by \citeN{be_2009}, since they compared and combined an ANN with different types of GARCH models (GARCH, EGARCH, GJR-GARCH, TGARCH, NGARCH, SAGARCH, PGARCH, APGARCH and NPGARCH). In addition to the above-mentioned researches, this type of hybrid models has been broadly used in other papers. \citeN{be_2014} proposed a MS-GARCH with an ANN to improve the forecasting accuracy, \citeN{bi_2011} combined a radial basis function with an EGARCH to model stocks returns of an Indonesian bank and \citeN{ap_2016} developed an ANN model as an extension of a GJR-GARCH to forecast the market returns of six European emerging markets. GARCH-based models have been also combined with ANNs to predict the volatility in commodity markets, such as gold (\citeNP{km_2015}) or oil (\citeNP{km_2016}). In this last case, the hybrid model included financial variables to improve the forecasts. This strategy can also be found in \citeN{kh_2017}. \citeN{kw_2018} propose a hybrid model that combines a LSTM with various GARCH-type models to forecast the volatility of KOSPI index. A refinement of this model can be found in \citeN{bk_2018}. It should be mentioned that these models can be generated in both directions: some outputs of a GARCH model can be used as input of an ANN and vice versa (\shortciteNP{lu_2016}). Finally, it should be noted that hybridisation can not only be made with ANN. \shortciteN{pmc_2018} proposed a structure combining traditional GARCH-models with Support Vector Machine (SVM) (\citeNP{Cortes_1995}).\\

The research carried out along this paper develops a volatility forecasting model that consists of two different levels which is based on stacking algorithms methodology (\shortciteNP{Hastie_2009}) and statistical models of the Machine Learning family. Random Forest (RF) (\citeNP{Breiman_2001}), Gradient Boosting (GB) with regression trees (\citeNP{Friedman_2000}) and Support Vector Machine (SVM) (\citeNP{Cortes_1995}) are used in the first level, while an ANN (\citeNP{Mcculloch_1943}) is incorporated within the second level of the stacked model (Stacked-ANN) in order to generate the volatility forecasts. A different two-level approach can be found in \citeN{km_2018}. They use an ANN-GARCH model with a pre-processing based on principal components analysis to reduce the number of inputs employed in their network. In contrast to the hybrid models defined previously, the proposed model is merging the results arising from other machine learning algorithms which are free of some theoretical assumptions like the use of a predefined distribution for the underlying asset returns or the constant level of unconditional variance. Because of this and with the aim to build a more flexible model, the GARCH-based models are not present in the Stacked-ANN architecture. The proposed model relies completely on the predictions made by machine learning algorithms and market data. Additionally, in the case of the Stacked-ANN the final forecasts made by the first level algorithms are directly used as inputs within the ANN while, in most of the hybrid models discussed in the previous paragraphs, sections of the GARCH-based models are inserted separately in the ANN. \\

The rest of the paper proceeds as follows: Section \ref{Previous} presents the set of volatility forecasting models used for comparison purposes. Furthermore, the risk measures and tests used to validate the results are discussed. In Section \ref{stack} the theoretical background and architecture of the volatility forecasting model based on stacking algorithms (Stacked-ANN) are explained. The empirical results of the different forecasting models are shown in Section \ref{resul}, where the accuracy and the risk measures arising from the proposed model are compared with results obtained by the methodologies explained in Section \ref{Previous}. Finally, Section \ref{conc} presents the main conclusions of the results and comparisons carried out along Section \ref{resul}.

\section{Benchmark models, risk measurements and statistical tests}
\label{Previous}
As stated above, this section is focused on explaining the benchmark models and the tests used to back-test the risk measurements. Thus, the first paragraphs are dedicated to ANN, ANN-GARCH, ANN-EGARCH and Heston Model, while the end of this section is focused on the risk measurements and tests performed to validate and compare the results of the benchmark models with the one proposed in Section \ref{stack}.\\

The first benchmark model is a feed-forward ANN. Following the notation provided by \citeN{Bishop_2006} and assuming that the algorithm has two hidden layers, the model would be defined by the following expression:
\begin{align}
\hat{\sigma}_{t+1}=h^{(3)} \left( \sum_{k=1}^T w_{p,k}^{(3)} h^{(2)} \left( \sum_{j=1}^M w_{k,j}^{(2)} h^{(1)} 
 \left( \sum_{i=1}^D w_{j,i}^{(1)} x_i + w_{j,0}^{(1)} \right) + w_{k,0}^{(2)} \right) + w_{p,0}^{(3)} \right)
\end{align}
Where \(h^{(n)}\) is the activation function associated with the layer \(n\), \(w_{z,v}^{(n)}\) is the \textit{v-th} weight associated with the neuron \(z\) inside the layer $n$ and \(x_i\) refers to the \(i\) input variable of database comprised by the explicative variables selected by the analyst.\\

The second benchmark model is an ANN-GARCH(\textit{p},\textit{q}). As briefly introduced in Section \ref{introd}, the aim of this hybrid model is to combine the GARCH(\textit{p},\textit{q}) estimates with other input variables by using an ANN, which is a more flexible model than GARCH(\textit{p},\textit{q}). Therefore, before starting with the fitting of the ANN, the parameters of the GARCH(\textit{p},\textit{q}) model need to be estimated:
\begin{equation}
\hat{\sigma}_t^2 =  \omega + \sum_{i=1}^{q}{\alpha_i r_{t-i}^2} + \sum_{i=1}^{p}{\beta_i \sigma_{t-i}^2}  \qquad  {\rm/} \quad \hat{r}_t =  \hat{\sigma}_t \epsilon_t 
\end{equation}
In this formulation \(\omega\), \(\alpha_i\) and \(\beta_i\) are the parameters to be estimated, while \(r_t\) and \(\sigma_t^2\) refer to the return and volatility respectively. The returns distribution is determined by the distribution selected for \(\epsilon_t\). If a standardize normal or standardize Student's t-distribution is selected, then the returns generated by the model follow a conditional normal (CND) or conditional t-distribution (CTD) respectively (\shortciteNP{Bauwens_2012}). Once the GARCH(\textit{p},\textit{q}) parameters are estimated, \(\sum_{i=1}^{q}{\alpha_i r_{t-1}^2}\) and \(\sum_{i=1}^{p}{\beta_i \sigma_{t-1}^2}\) can be computed and used as input (together with the rest of explicative variables) within the ANN.\\

The third benchmark model is an ANN-EGARCH. The architecture of this model and the previous one can be considered the same with the unique difference that the first step consists of fitting an EGARCH(\textit{p},\textit{q}) instead of a GARCH(\textit{p},\textit{q}) model. The EGARCH(\textit{p},\textit{q}) can be defined as follows (\citeNP{Nelson_1991}):
\begin{align}
&\log{\hat{\sigma}_t^2} =  \omega + \sum_{i=1}^{p}{\alpha_i \log{\hat{\sigma}_{t-i}^2}} + \sum_{i=1}^{q}{(\beta_i \epsilon_{t-i}+\gamma_i(\mid \epsilon_{t-i} \mid-E\mid \epsilon_{t-i} \mid))}
\end{align}
Once the EGARCH is fitted, the following terms can be calculated and used as input within the ANN together with the rest of the explicative variables selected by the analyst:
\begin{align}
\sum_{i=1}^{p}{\alpha_i \log{\hat{\sigma}_{t-i}^2}}&
&\sum_{i=1}^{q}{\beta_i \epsilon_{t-i}}&
&\sum_{i=1}^{q}{\gamma_i(\mid \epsilon_{t-i} \mid-E\mid \epsilon_{t-i} \mid)}
\end{align}
The last benchmark is the \citeN{Heston_1993} Model. Even though this approach belongs to the stochastic family and the proposed one to the Machine Learning one, this model is going to be used as benchmark during this paper as this process is the most widely used within the family of the stochastic volatility models. It assumes that changes in stock prices through the time ($dX_t$) follow a Brownian diffusion process:
\begin{align}
dX_t &= \mu X_t dt + \sqrt{\sigma_t^2} X_t dB_t        
\end{align}
Where \( B_t \sim \mathcal{N}(0,\sigma_t^2t) \). Therefore, if volatility follows an Ornstein-Uhlenbeck process, the changes in this variable are defined by the following expression: 
\begin{align}
d\sigma_t^2 &= \theta (\upsilon - \sigma_t^2) dt + \delta \sigma_t dB_t^{*}
\end{align}
where \(\upsilon\) is the long term volatility, \(\theta\) is the rate of return to \(\upsilon\), \(\delta\) is the volatility of \(\sigma_t^2\) and \(B_t^{*}\) is a Wiener process that has a correlation of \(\rho\) with \(B_t\).\\

Once the four benchmark models have been explained, the section focuses on the risk measurements. As stated before, volatility plays a key role in market risk assessment. Therefore, the models will not be only compared in terms of accuracy, but the risk measurements arising from every volatility model are going to be tested. For this purpose, VaR and CVaR have been selected as risk measures. Even though VaR is probably the most used metric due to its simplicity and easy interpretation, CVaR has been also included as it is considered to be a coherent risk measure (\shortciteNP{Artzner_1999}). Consequently, for every volatility model the aforementioned risk measures are going to be computed and validated by means of the following tests:
\begin{itemize}
\item \citeN{Kupiec_1995} introduced a test in order to check if the number of VaR excesses are align with the level of confidence selected.
\item An extension of the previous test was developed by \shortciteN{Christoffersen_1997}. The aim of this test is to validate that VaR excesses are independent, identically distributed and in line with the selected level of confidence.
\item \citeN{Acerbi_2014} developed a test (AS1) to assess the appropriateness of the CVaR based on the assumption that VaR has been already tested and considered to be correct from a statistical point of view. The test is inspired by the following equation:
\begin{align}
E \left[ \frac{r_t}{CVaR_{\alpha , t}} +1 \bigg| r_t + VaR_{\alpha , t} < 0\right]=0
\end{align}
As VaR needs to be previously validated, the result of this test has to be assessed together with the two aforementioned tests.
\item In addition to the previous test, \citeN{Acerbi_2014} introduced another method (AS2) to validate the CVaR without making any assumption about the appropriateness of the VaR. To do so, this test tries to check a CVaR expression that is not conditioned by the correctness of a previous VaR estimate. 
\end{itemize}
Before beginning with the Stacked-ANN architecture, it is worth noticing that the two first tests are parametric while the two last are non-parametric so, for further details about how to compute the statistics and their distributions please refer to aforementioned papers.

\section{Stacked model}
\label{stack}
\begin{figure}[hbt!]
\begin{center}
\caption{Stacked-ANN model structure}
\includegraphics[width=0.99\textwidth]{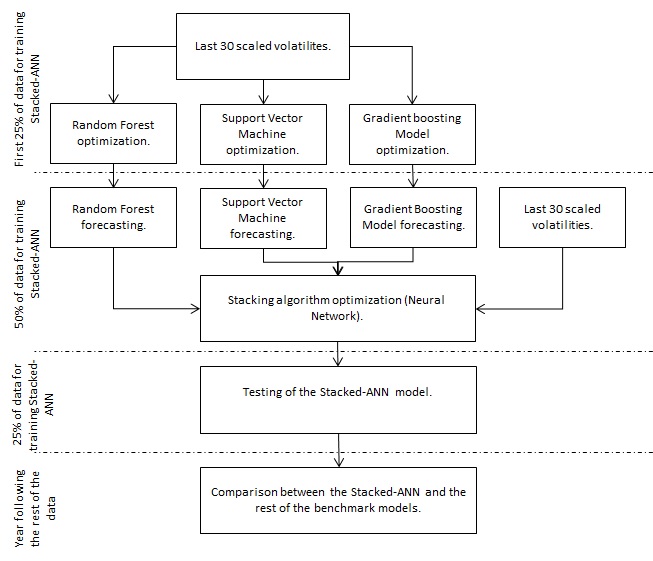}
\label{fig:Fig 1}
\end{center}
\end{figure}

This section has been divided in several sub-sections in order to explain sequentially the proposed volatility forecasting model. As the Stacked-ANN model is composed by two different levels, the two first sub-sections are dedicated to the input data and the algorithms within the first level of the Stacked-ANN model, while the third and forth sub-sections are focused on the data required to generate the stacking procedure and the details of the ANN fitted with the aforementioned information. (Figure \ref{fig:Fig 1} explains briefly the process followed to estimate and test the Stacked-ANN model)

\subsection{First level: Input data}
\label{1Inputs}
The first step is concerned with the creation of the database containing the volatility proxy to be used as a response and the explanatory variables selected to fit the algorithms. As the aim of the study is to predict future volatilities, the True Realized Volatility (hereinafter TRV) is going to be used as response variable (\citeNP{Roh_2006}):
\begin{align}
TRV_t &= \sqrt{ \frac{1}{n} \sum_{i=1}^{n} (r_{t+i-1}-\widehat{r}_t)^2} 
\end{align}
Where \(\widehat{r}_t= \sum_{i=n}^{n}(r_{t+i-1})/n\) and \(n=5\). The window has been selected to be large enough to compute a stable TRV and small enough to avoid, as much as possible, mixing different volatility regimes.\\

The variables given to the first level algorithms to forecast the TRV  are the last 30 volatilities computed with returns already observed in the market:
\begin{align}
V_t &= \sqrt{ \frac{1}{n} \sum_{i=0}^{n-1} (r_{t-n+i}-\widehat{r}_t)^2} 
\end{align}
Where \(\widehat{r}_t= \sum_{i=0}^{n-1}(r_{t-n+i})/n\) and \(n=5\). Only the last 30 volatilities have been selected because the correlations between previous volatilities and the TRV are residual and therefore their explanatory power is considered to be non-significant. The historical data to compute all the aforementioned variables is obtained by using the \texttt{quantmod} (\citeNP{Ryan_2017}) package from the R project (\citeNP{RCT_2017}) and, as suggested by \shortciteN{Hastie_2009}, they will be scaled to the range \([0,1]\) to improve the training of the algorithms.\\

Before beginning with the section related with the algorithms included within the first level, it is important to mention that the first 25\% of the data is used to fit the first level algorithms, the next 50\% is dedicated to the ANN estimation and the last 25\% is the test set. The comparison of the benchmark models with the proposed one in terms of accuracy and risk measurement will be made with a different set of data containing the information of the following year (e.g. if data from 2000 to 2007 is used to train and test the Stacked-ANN model, the out of sample data selected for comparison purposes would be market movements happened during 2008).

\subsection{First level: Individual algorithms}
\label{1Alg}
The methods applied to optimize the hyper-parameters of the algorithms within the first level of the Stacked-ANN architecture are introduced below:
\begin{itemize}
\item Minimization of the Mean Square Error (hereinafter, MMSE) for the whole database to train the first level algorithms.
\item Circular Block Bootstrap (CBB). This method (\citeNP{Politis_1991}) generates new samples by selecting random blocks from the original database. The length of these blocks is fixed and the procedure to calculate it was introduced by \citeN{Politis_2004} and \shortciteN{Patton_2009}. CBB can only be applied to stationary time series.
\item Stationary Bootstrap (hereinafter, SB) (\citeNP{Politis_1994}). Similar to the case of CBB, this method can only be used with stationary time series. However, the difference with the former method is that the length of the blocks instead of being fixed, it is randomly selected with a certain average that can be calculated using different approaches (see \citeNP{Politis_2004} and \shortciteNP{Patton_2009}).
\item Maximum Entropy Bootstrap (hereinafter, MEB) (\citeNP{Vinod_2006} and \citeNP{Vinod_2009}). Unlike the two previous approaches, stationarity is not required as the new samples are obtained from the maximum entropy distribution of the original time series.
\item H Cross-Validation (HCV). This method introduced by \citeN{Chu_1991} tries to avoid the effect of the correlation that can exist between the response and the explanatory variables while dealing with time series by eliminating {h} data points between them. The bandwidth selection is obtained minimizing the absolute autocorrelation between the response and explanatory variables, with a maximum width of 100 days.
\end{itemize}
The optimum hyper-parameters combination of each one of the five previous methods is obtained by applying grid search. Then, these combinations are tested against data out of sample (the following 50\% of the database) to choose the most accurate option for fitting the algorithm.\\

As stated before, the first level of the stacked model architecture is composed by three algorithms: Random Forest (RF) (\citeNP{Breiman_2001}), Gradient Boosting with regression trees (GB) (\citeNP{Friedman_2000}) and Support Vector Machine (SVM) (\citeNP{Cortes_1995}).\\

\subsection{Second level: Input data}
\label{2Inputs}
As explained in Section \ref{1Inputs}, the first 25\% percent of data is dedicated to fit the first level algorithms while the following 50\% and 25\% are used for fitting the ANN and testing the results respectively. The explanatory variables given to the ANN are:
\begin{itemize}
\item As with the first level algorithms, the last 30 volatilities \((V_t,V_{t-1},...,V_{t-29})\) scaled to the range \([0,1]\).
\item The True Realized Volatility forecasts made by the first level algorithms: Random forest (\(\widehat{TRV}_{t, RF}\)), Gradient boosting (\(\widehat{TRV}_{t, GB}\)) and Support Vector Machine (\(\widehat{TRV}_{t,SVM}\)).
\end{itemize}
The response variable is the \(TRV_t\) as defined in Section \ref{1Inputs}.
\subsection{Second level: Stacking algorithm}
\label{2Alg}
As stated previously, the last step of the Stacked-ANN model is the fitting of the ANN, which is the algorithm stacking the forecasts made by the RF, GB and SVM. Before starting with the details of the ANN architecture, notice that the methods and procedures related to the hyper-parameters optimization are the same as the first level algorithms: Grid search in combination with the methods explained in Section \ref{1Alg} and final hyper-parameters decision based on the out of sample error (last 25\% of the database).\\

Below, the main characteristics and details of the stacking algorithm are presented:
\begin{itemize}
\item The feed-forward ANN has two hidden layers with 20 and 10 neurons respectively. The sigmoid activation function has been selected for all the neurons within the hidden layers while the linear activation function has been used in the output layer, which is comprised by one neuron.
\item The optimization algorithm selected is Adaptive Moment Estimation (ADAM), which was created by \shortciteN{dk_2014}. This method consists in a progressive adaptation of the initial learning rate, taking into consideration current and previous gradients. The default calibration proposed by the authors is applied: \(\beta_1=0.9\) and \(\beta_2=0.999\).
\item The number of epochs are 10,000 and the batch size is equal to the length of the data used for training the ANN.
\item The backward pass calculations are done according to the selection of root mean squared error as a loss function.
\item As indicated in Section \ref{1Inputs}, the 50\% of the information is selected for training the ANN while the following 25\% of the data is the test set. Note that the first 25\% of the data is used to fit the first level algorithms.
\item The parameter adjusting the level of L2 regularization \((\phi)\) and the initial learning rate \(\lambda\) used within ADAM are the hyper-parameters to be optimized during the estimation process.
\end{itemize}
Taking into consideration all the above-mentioned details, the \(TRV_t\) forecasted by the Stacked-ANN model (S-ANN) is obtained by means of the following expression:
\begin{align}
\begin{split}
&\widehat{TRV}_{t,S-ANN}=\widehat{f}(\widehat{TRV}_{t, RF},\widehat{TRV}_{t, GB},\widehat{TRV}_{t, SVM},V_t,V_{t-1},...,V_{t-29})=\\
&=h^{(3)} \left( \sum_{k=1}^{10} w_{1,k}^{(3)} h^{(2)} \left( \sum_{j=1}^{20} w_{k,j}^{(2)} h^{(1)} 
 \left( \sum_{i=1}^{33} w_{j,i}^{(1)} x_i + w_{j,0}^{(1)} \right) + w_{k,0}^{(2)} \right) + w_{1,0}^{(3)} \right)
\end{split}
\end{align}
As explained in Section \ref{2Inputs}, \(x_i\) are the last 30 volatilities scaled to the range \([0,1]\) and the forecasts made by the first level algorithms.

\newpage
\section{Results}
\label{resul}
During this section, the data used in the empirical analysis, the fitting process and the final comparison between the Stacked-ANN and the benchmark models are shown.
\subsection{Data}
\label{ResultsData}
In order to analyse the models under different market conditions, the algorithms have been trained and tested five different times with the S\&P 500 volatilities observed in the following periods: 2000-2007, 2001-2008, 2002-2009, 2009-2016 and 2010-2017. As stated in Section \ref{1Inputs}, during the training and testing of the models the first 25\% of the periods selected is dedicated to fit the first level algorithms, the next 50\% is used to optimize the ANN while the last 25\% is reserved for testing purposes. The year after the aforementioned periods (2008, 2009, 2010, 2017 and 2018 respectively for each period) has been used to compare the out of sample results of the Stacked-ANN with the benchmark models. The first three data-sets have been selected in order to analyse the performance of the models during the years after the financial crisis, when the markets where dominated by a high volatile regime. Although the years influenced by the financial crisis are valuable to test the accuracy of the volatility forecasting models, the two last data-sets have been selected in order to analyse the models performance with the most recent data. Additionally, the lower level of volatility during the last periods, especially in 2017, allows to assess the robustness of the models by analysing them in different market conditions. In order to support the explanations given during this paragraph, Table \ref{VolStats} summarizes the moments of the TRV during the different periods selected to compare the models:\\

\begin{table}[h]
  \begin{center}
    \caption{True Realised Volatility statistics}
    \label{VolStats}
    \begin{tabular}{ L{3cm} C{2cm} C{2cm} C{2cm} C{2cm}}
      \hline
      Period       &   Mean         &   STD          &  Skewness     &   Kurtosis    \\
      \hline
      Year 2008    &   0.022        &   0.016        &   1.510       &   4.519       \\ 
      Year 2009    &   0.015        &   0.008        &   0.853       &   3.248       \\ 
      Year 2010    &   0.010        &   0.006        &   0.854       &   3.736       \\ 
      Year 2017    &   0.004        &   0.002        &   0.911       &   3.369       \\ 
      Year 2018    &   0.009        &   0.006        &   1.406       &   4.702       \\ 
      \hline
     \multicolumn{2}{l}{\emph{Source}: own elaboration}    
    \end{tabular}
  \end{center}
\end{table}

In addition, the Kolmogorov-Smirnov test has been applied sequentially to the TRV in order to assess statistically if the behaviour of the volatility changes over the different periods. As 2008 is the year when the most extreme events related with crisis happened and the market changed from a low to a high volatile regime, the skewness and mean of that year volatility is higher than the one related with 2009. Because of that, the aforementioned test reveals that the volatility of 2008 and 2009 do not belong to the same distribution (\(KS_{p-value}=0.001\)). However, when comparing the volatility of 2009 with the 2010 one, the test indicates that they come from the same distribution (\(KS_{p-value}=0.690\)). Even though the volatility follows a downward trend, both years are heavily conditioned by the events occurred during 2008 and therefore the test accepts the hypothesis that volatilities belong to the same distribution. Finally, the pair comprised by the volatilities of 2017 and 2018 shows an upward trend. Nevertheless, this increase is not big enough to reject the hypothesis that they come from the same distribution (\(KS_{p-value}=0.167\)).\\

The use of some of the methods proposed in Section~\ref{1Alg} requires the time series to be stationary. Therefore, before using block bootstrap it has been checked if historical volatility satisfies this property by applying the Augmented Dickey-Fuller test (\citeN{ADF_1979}) to the different data-sets dedicated to fit the algorithms within the first and second level. The results are shown in Table~\ref{ADF}:\\
 \begin{table}[h]
  \begin{center}
    \caption{Augmented Dickey-Fuller Test}
    \label{ADF}
    \begin{tabular}{l c c}
      \hline
                     & ADF statistic: Data for           & ADF statistic: Data for        \\
      Period         & training 1st level                & training 2nd level             \\
      \hline
      (2000-2007)    &   -6.61                           & -6.41                          \\ 
      (2000-2008)    &   -6.13                           & -7.91                          \\ 
      (2000-2009)    &   -5.25                           & -7.57                          \\  
      (2009-2016)    &   -4.72                           & -7.27                          \\
      (2010-2017)    &   -4.58                           & -8.82                          \\ 
      \hline
     {\emph{Source}: own elaboration}    
    \end{tabular}
  \end{center}
\end{table}

As the critical values are \(-2.63\) and \(-3.43\) with a probability of 5\% and 1\% respectively, it can be concluded that the data meet the requirements imposed by CBB and SB methods.\\

Previously to the fitting of the algorithms, the parameters needed for the different bootstrap and cross validation methods are obtained by means of the methodologies presented in Section \ref{1Alg}. As the Stacked-ANN architecture is comprised by two different levels, the length of blocks for CBB, the average of the blocks for SB and the distance, $h$, to be used within the HCV method are obtained for both, the data-set to fit first level algorithms and the one dedicated to the second level. Table~\ref{param} summarizes the former parameters and it shows non-significant changes over time for the different periods and levels:\\

 \begin{table}[h]
  \begin{center}
    \caption{Calibration of the elements for bootstrap and CV}
    \label{param}
    \begin{tabular}{l l c c}
      \hline
                         &                    & Data for training        & Data for training     \\
      Method             & Period             & 1st level algorithms     & 2nd level algorithm   \\
      \hline
      CBB Block          & (2000-2007)        &   28                     & 63                    \\ 
      CBB Block          & (2001-2008)        &   36                     & 58                    \\ 
      CBB Block          & (2002-2009)        &   40                     & 56                    \\ 
      CBB Block          & (2009-2016)        &   39                     & 58                    \\ 
      CBB Block          & (2010-2017)        &   38                     & 30                    \\ 
      SB Block average   & (2000-2007)        &   25                     & 55                    \\ 
      SB Block average   & (2001-2008)        &   32                     & 51                    \\ 
      SB Block average   & (2002-2009)        &   35                     & 49                    \\ 
      SB Block average   & (2009-2016)        &   34                     & 51                    \\ 
      SB Block average   & (2010-2017)        &   33                     & 27                    \\ 
      HCV length         & (2000-2007)        &   26                     & 31                    \\ 
      HCV length         & (2001-2008)        &   31                     & 51                    \\ 
      HCV length         & (2002-2009)        &   31                     & 40                    \\ 
      HCV length         & (2009-2016)        &   32                     & 55                    \\ 
      HCV length         & (2010-2017)        &   35                     & 27                    \\ 
      \hline
     {\emph{Source}: own elaboration}    
    \end{tabular}
  \end{center}
\end{table}

\newpage
\subsection{Fitting of the Stacked-ANN model}
\label{ResultsFit}
As explained in Section \ref{1Alg}, different approaches have been followed to find the optimum hyper-parameter combination. Table \ref{Optimum} shows the methods that minimize the out of sample error per each algorithm and period:\\
\begin{table}[h]
  \begin{center}
    \caption{Methods optimizing OOS error}
    \label{Optimum}
    \begin{tabular}{l c c c c}
      \hline
                  & Stacking           &                  &   Gradient    &  Support              \\
      Period      & Algorithm (ANN)    &  Random Forest   &   Boosting    &  Vector Machine       \\
      \hline
      (2000-2007) &   ME               &    SB            & CBB           &   SB                  \\ 
      (2001-2008) &   CBB              &    CBB           & CBB           &   SB                  \\ 
      (2002-2009) &   CBB              &    CBB           & CBB           &   CBB                 \\  
      (2009-2016) &   HCV              &    HCV           & HCV           &   SB                  \\ 
      (2010-2017) &   SB               &    CBB           & SB            &   SB                  \\ 
      \hline
     \multicolumn{2}{l}{\emph{Source}: own elaboration}    
    \end{tabular}
  \end{center}
\end{table}

Regardless of the period, the empirical results suggest that CBB and SB outperform the rest of the methods. These outcomes are expected as these two methods based on re-sampling blocks from the original database are specifically prepared to work with stationary time series. Table \ref{Hyper} summarizes the hyper-parameters suggested by the methods shown in Table \ref{Optimum}:\\
\begin{table}[h]
  \begin{center}
    \caption{Final hyper-parameters}
    \label{Hyper}
    \begin{tabular}{l c c c c}
      \hline
                  & Stacking          &                 &   Gradient     &  Support              \\
      Period      & Algorithm (ANN)   &  Random Forest  &   Boosting     &  Vector Machine       \\
      \hline
      (2000-2007) & $\phi=0$          & $N=10$          & $B=1479$       & $\gamma=0.0001$       \\ 
                  & $\lambda=0.0033$  & $Obs=24$        & $\lambda=0.003$ & $\epsilon=0.45$       \\ 
      \hline
      (2001-2008) & $\phi=0.01$       & $N=10$          & $B=3000$       & $\gamma=0.0001$       \\ 
                  & $\lambda=0.0059$  & $Obs=107$       & $\lambda=0.001$& $\epsilon=0.55$       \\ 
      \hline
      (2002-2009) & $\phi=0$          & $N=1$           & $B=3583$       & $\gamma=0.0004$       \\ 
                  & $\lambda=0.0136$  & $Obs=37$        & $\lambda=0.001$& $\epsilon=0.17$       \\ 
      \hline
      (2009-2016) & $\phi=0.02$       & $N=30$          & $B=1000$       & $\gamma=0.0002$       \\ 
                  & $\lambda=0.085$   & $Obs=118$       & $\lambda=0.009$& $\epsilon=0.13$       \\       
      \hline
      (2010-2017) & $\phi=0.01$       & $N=7$           & $B=1000$       & $\gamma=0.0001$       \\ 
                  & $\lambda=0.011$   & $Obs=175$       & $\lambda=0.003$& $\epsilon=0.54$       \\ 
                  
      \hline
     \multicolumn{2}{l}{\emph{Source}: own elaboration}    
    \end{tabular}
  \end{center}
\end{table}

\newpage
Where \(\lambda\) is the learning rate of the ANN and GB, \(\phi\) the parameter adjusting the level of L2 regularization of the ANN, \(B\) the number of iterations performed while fitting the GB, \(N\) the number of variables randomly selected by the RF and \(Obs\) the minimum number of observations to be kept in the terminal nodes of every fitted tree within the RF architecture. Finally, \(\gamma\) refers to the parameter included within the radial basis function kernel (the lower the parameter, the higher the non-linearity) and \(\epsilon\) defines the threshold where the error begins to be penalized by the SVM.

\subsection{Comparison against benchmark models}
\label{ResultsCompare}
Once the Stacked-ANN is fitted, its performance is compared with the benchmark models explained in Section \ref{Previous} (ANN, ANN-GARCH(1,1), ANN-EGARCH(1,1) and Heston Model). Before beginning with the comparisons, the three following remarks about the benchmark models have to be done:
\begin{itemize}
\item Due to the nature of the Heston Model, 20,000 simulations per each day have been computed and the daily average of them has been taken to assess its accuracy. 
\item The GARCH(1,1) and EGARCH(1,1) (included in the ANN-GARCH(1,1) and ANN-EGARCH(1,1) architecture respectively) have been estimated assuming Student-t innovations.
\item The fitting procedure and architecture of the ANNs included within ANN-GARCH(1,1), ANN-EGARCH(1,1) and ANN models are the same as the ones explained for the Stacked-ANN (see Section \ref{2Alg}).
\end{itemize}
\begin{table}[h]
  \begin{center}
    \caption{Accuracy analysis}
    \label{Accuracy}
    \begin{tabular}{L{3cm} C{1.65cm} C{1.65cm} C{1.65cm} C{1.65cm} C{1.65cm}}
      \hline
                  & RMSE:          & RMSE:          & RMSE:         & RMSE:        & RMSE:         \\
      Model       & 2008           & 2009           & 2010          & 2017         & 2018          \\
      \hline
      Stacked-ANN &   0.01192      &   0.00534      &   0.00494     &   0.00254    &   0.00544     \\ 
      ANN-EGARCH  &   0.01332      &   0.00588      &   0.00537     &   0.00276    &   0.00571     \\ 
      ANN-GARCH   &   0.01335      &   0.00584      &   0.00539     &   0.00263    &   0.00575     \\ 
      Heston      &   0.02066      &   0.00714      &   0.00547     &   0.00359    &   0.00610     \\ 
      ANN         &   0.01526      &   0.00615      &   0.00541     &   0.00274    &   0.00590     \\ 
      \hline
     \multicolumn{2}{l}{\emph{Source}: own elaboration}    
    \end{tabular}
  \end{center}
\end{table}

Table \ref{Accuracy} shows the out of sample error of the different periods selected to compare the performance and robustness of the Stacked-ANN with the benchmark models. The results shown in this table suggest the following conclusions:
\begin{itemize}
\item Regardless of the period, the Stacked-ANN outperforms other hybrid models based on auto-regressive methodologies like ANN-GARCH and ANN-EGARCH. In relative terms, minor deviations are observed between the different periods. 
\item All the hybridized models tend to outperform the pure ANN model.
\item As expected due to the extremely high volatilities observed during the financial crisis, the results show that, regardless of the model, 2008 forecasts are less accurate. All the models minimize their error rate in the year with the lowest level volatility, 2017.
\item The forecasts made by the Heston Model tend to be the less accurate due to the non-predictive nature of this model.
\end{itemize}
In addition to the above-mentioned analysis, the risk measures obtained by using each one of the volatility models are tested. In order to do so, a returns distribution is selected for each one of the forecasting volatility methods. As described in Section \ref{Previous}, Heston Model requires the changes in stock prices to follow a Brownian diffusion process. Nevertheless, for the rest of the benchmark models and the Stacked-ANN (which are free of assumptions about the returns) a Student t-distribution has been combined with the different volatility forecasts. This assumption about Student t-distribution has been selected when possible as returns tend to be leptokurtic and heavier-tailed than Normal distribution (\shortciteNP{McNeil_2015}).\\

Before analysing the results of the tests presented in Section \ref{Previous}, it is worth mentioning that the level of confidence (99\%) and number of days (10) selected are based on the ones set by Basel Directive, whose aim is to monitor, amongst others, the market risk. Table \ref{RiskTable} shows the p-value of the tests dedicated to VaR (Kupiec and Christoffersen) and CVaR (AS1 and AS2). If a 95\% is set as confidence level, Stacked-ANN in combination with Student t-distribution is the only model that produces an appropriate p-value for Kupiec, AS1 and AS2 tests in every period under analysis. All the models show difficulties to produce a p-value higher or equal than \(0.05\) for the Christoffersen test because VaR exceedances tend to happen in a short period of time instead of spread over the period analysed. It is worth mentioning that the hybrid models taken as benchmark (ANN-EGARCH and ANN-GARCH) also fail in producing an appropriate value for the Kupiec test in several periods while, as stated before, the proposed hybrid model (Stacked-ANN) pass the test for every period. Finally, Heston Model tends to produce less appropriate risk measures due to the distribution constrain mentioned previously.

\begin{table}[h]
  \begin{center}
    \caption{P-value of the VaR and CVaR tests}
    \label{RiskTable}
    \begin{tabular}{l c c c c c c}
      \hline
                  &          &  Period:       &  Period:       &  Period:       &  Period:       &  Period:       \\
      Model       & Test     &  2008          &  2009          &   2010         &    2017        &    2018        \\
      \hline
      Stacked-ANN & Kupiec   & 0.85           & 0.84           & 0.65           & 0.85           & 0.85           \\ 
                  & Christ.  & 0.01           & 0.79           & 0.02           & 0.01           & 0.01           \\ 
                  & AS1      & 0.66           & 0.85           & 0.61           & 0.90           & 0.91           \\ 
                  & AS2      & 0.56           & 0.63           & 0.36           & 0.67           & 0.69           \\ 
      \hline
      ANN-EGARCH  & Kupiec   & 0.12           & 0.12           & 0.84           & 0.03           & 0.03           \\ 
                  & Christ.  & 0.00           & 0.00           & 0.01           & 0.03           & 0.03           \\ 
                  & AS1      & 0.52           & 0.85           & 0.61           & 1.00           & 1.00           \\ 
                  & AS2      & 0.07           & 0.19           & 0.62           & 0.91           & 0.91           \\ 
      \hline
      ANN-GARCH   & Kupiec   & 0.12           & 0.03           & 0.01           & 0.03           & 0.03           \\ 
                  & Christ.  & 0.00           & 0.03           & 0.00           & 0.03           & 0.03           \\ 
                  & AS1      & 0.51           & 1.00           & 0.77           & 1.00           & 1.00           \\ 
                  & AS2      & 0.08           & 0.92           & 0.05           & 0.85           & 0.89           \\ 
      \hline
     Heston Model & Kupiec   & 0.00           & 0.00           & 0.65           & 0.03           & 0.00           \\ 
                  & Christ.  & 0.00           & 0.00           & 0.59           & 0.03           & 0.00           \\ 
                  & AS1      & 0.00           & 0.01           & 0.83           & 1.00           & 0.06           \\ 
                  & AS2      & 0.00           & 0.00           & 0.36           & 0.92           & 0.00           \\
      \hline
      ANN         & Kupiec   & 0.65           & 0.04           & 0.65           & 0.30           & 0.29           \\ 
                  & Christ.  & 0.02           & 0.00           & 0.00           & 0.00           & 0.00           \\ 
                  & AS1      & 0.24           & 0.86           & 0.59           & 0.81           & 0.00           \\ 
                  & AS2      & 0.29           & 0.11           & 0.35           & 0.24           & 0.00           \\ 

     \hline
     \multicolumn{2}{l}{\emph{Source}: own elaboration}    
    \end{tabular}
  \end{center}
\end{table}

\newpage
\section{Conclusions}
\label{conc}
This paper introduces a Stacked-ANN model based only on Machine Learning techniques with the aim to improve the accuracy of the volatility forecasts made by other hybrid models based on a combination of GARCH or EGARCH with ANNs. Its predictive power and performance has been tested in terms of RMSE, VaR and CVaR.\\

Two main results have to be pointed out. Firstly, the Stacked-ANN has been able to generate more accurate volatility forecasts than other models in a high volatile regime period like the one occurred after the Financial Crisis of 2007-2008. The models outperformed by the Stacked-ANN during that time lapse are other hybrid models like ANN-GARCH and ANN-EGARCH, the most widely used stochastic volatility theory (Heston Model) and a feed-forward ANN without any combination with other algorithms or statistical models. Notwithstanding the Stacked-ANN performance, it is observed for every model that the higher the volatility the lower the accuracy. In addition to this analysis, the Stacked-ANN has been tested with the most recent data (2017 and 2018) in order to check its performance in the current market conditions. As it occurred with the tests carried out during the financial crisis, the proposed architecture outperforms the benchmark models in terms of accuracy. The superior performance shown by the Stacked-ANN in periods with different levels of volatility are due to the model flexibility. In contrast with ANN-GARCH or ANN-EGARCH, the inputs introduced in the ANN stacked model do not follow any theoretical assumption about the returns distribution or volatility. As explained throughout Section \ref{stack}, the architecture proposed uses previous volatilities and forecasts made by a random forest, gradient boosting with regression trees and support vector machine as inputs. Before beginning with the second point of the conclusion, it is worth mentioning that it has been empirically demonstrated that block bootstrap methods are of special interest when fitting algorithms to volatility as these procedures are especially prepared to work with stationary time series.\\

Secondly, the forecasts made by the volatility models have been combined with a certain distribution in order to compute the VaR and CVaR for all the different periods analysed. The distribution selected has been the Student's t-distribution for every model with the exception of the Heston Model which requires changes in asset prices to follow a Brownian diffusion process. The empirical results demonstrated that only the Stacked-ANN model is able to produce an appropriate p-value for Kupiec, AS1 and AS2 tests in every period under analysis, including those ones related with the financial crisis.\\

The aforementioned flexibility and predictive power of the Stacked-ANN compared with other volatility models suggest to develop further investigations about the implications of using this model for derivative valuation purposes. As the price of these instruments is closely related to the volatility of the underlying assets, further researches should be done in order to compare the implied volatilities observed in the market with the ones arising from the proposed model. If the volatility measured by the Stacked-ANN is more accurate than market expectations, it would be possible to identify under and overvalued derivatives.\\

\bibliography{Final1}
\bibliographystyle{chicago}
\end{document}